# Generation of narrow-band polarization-entangled photon pairs at a rubidium D1 line[*]

Long Tian (田龙), Shu-Jing Li (李淑静)[†], Hao-Xiang Yuan (原浩翔)

and Hai Wang (王海)[†]

*The State Key Laboratory of Quantum Optics and Quantum Optics Devices,*

*Collaborative Innovation Center of Extreme Optics, Institute of Opto-Electronics,*

*Shanxi University, Taiyuan, 030006, People's Republic of China*

Using the process of cavity-enhanced spontaneous parametric down-conversion (SPDC), we generate a narrow-band polarization-entangled photon pair resonant on the rubidium (Rb) D1 line (795 nm). The degenerate single-mode photon pair is selected by multiple temperature controlled etalons. The linewidth of generated polarization-entangled photon pairs is 15 MHz which matches the typical atomic memory bandwidth. The measured Bell parameter for the polarization-entangled photons S=2.73±0.04 which violates the Bell-CHSH inequality by ~18 standard deviations. The presented entangled photon pair source could be utilized in quantum communication and quantum computing based on quantum memories in atomic ensemble.

---

[*] Project supported by the National Basic Research Program of China (Grant No. 2016YFA0301402), the National Natural Science Foundation of China (Grant Nos. 11475109, 11274211, and 60821004) , and the Program for Sanjin Scholars of Shanxi Province of China.

[†]Corresponding authors. Email: lishujing@sxu.edu.cn and wanghai@sxu.edu.cn

# 1. Introduction

Storage of entangled photons is a key component for long-distance communication and linear quantum computation [1-6]. Spontaneous parametric down-conversion (SPDC) in nonlinear crystals is the most widely used method to generate entangled photons [7-10]. However, the entangled photons from single-pass SPDC have a very broad linewidth on the order of THz [8-9] which are unfeasible to be stored in the most matter nodes [2, 11-14]. For instance, the typical bandwidths of the atomic quantum memories range from MHz to GHz [14-18]. In order to efficiently store the entangled photons in atomic memories, we need to reduce the linewidth of entangled photons from SPDC.

By means of optical filters, the linewidths of the photons from SPDC have been reduced to GHz even MHz order and the storages of the entangled photons have been demonstrated in solid-state [11-12] and atomic ensembles [14]. The drawback of the filtering scheme is that the count rate of photons is inevitably decreased. For solving this problem, cavity-enhanced SPDC has been developed to prepare the narrow-band photon pairs [19]. By putting the nonlinear crystal inside a cavity, the linewidth of generated photon pairs is limited by the cavity linewidth and the generation probability for the down-converted photons whose frequency matches the cavity mode will be enhanced greatly. Significant progress has been made in this regard [19-25]. Recently, the generation of single-mode narrow-band polarization entangled source has been reported through cavity-enhanced SPDC [25], and the storage of such entangled photons has been demonstrated in a cold atomic ensemble [26].

In this paper, we also demonstrate a narrow-band polarization-entangled photon pairs at Rb $D_1$ line (795 nm). Based on cavity-enhanced SPDC, a pair of horizontally (H) and vertically (V) polarized photons is generated from a type-II periodically poled $KTiOPO_4$ (PPKTP) crystal placed in a cavity. Passing though multiple etalons, the non-degenerate correlated photon pairs from the cavity are filtered and the degenerate single-mode pairs are remained. The single-mode photon pair is split into two parts with a 50/50 non-polarizing beam splitter (NPBS). For the case that each output port has one photon, the photon pair will be in polarization entangled state in the post-selected manner. In contrast to the past experiment [25], in which, $|H\rangle$ and $|V\rangle$ modes of the photon pairs are separated on a polarizing beam splitter (PBS), and filtered by two different etalons, respectively, then combined on another PBS for generating entangled photon pairs, in our experiment, the $|H\rangle$ and $|V\rangle$ modes of the photon pairs are filtered by a set of etalons, and then sent to an 50/50 NPBS to generate entangled two polarization qubits. So our experimental setup is relative simple，which will benefit the practical application of entanglement source in quantum information process. The linewidth of polarization-entangled photon pairs here is 15 MHz, which matches the typical atomic memory bandwidth [2]. The measured Bell parameter for the polarization-entangled photons $S$=2.73±0.04 which violates the Bell-CHSH inequality by ~18 standard deviations.

## 2. Experimental setup

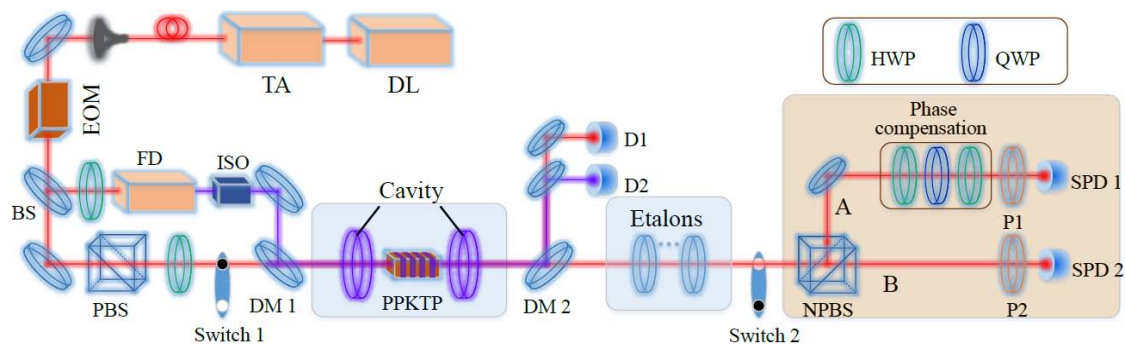

**Fig.1 (color online)** Experimental set-up. EOM: Electro-Optic modulator; FD: frequency doubler; ISO: Optical isolator; DM1-2: Dichroic mirrors; NPBS: Non-polarizing beam splitter; HWP: Half wave plate; QWP: Quarter wave plate; P1-2: Polarizers; D1-2: Photodetectors; SPD1-2: Single photon detectors; Switch1-2: Mechanical switches.

The experimental setup is shown in Fig. 1. A tapered amplifier (TA) is employed to amplify the output power of diode laser which is locked to the $D_1$ line of Rb, the corresponding wavelength is 795 nm. The beam from the TA is phase modulated by an electro-optic modulator. This modulation is used to lock a cavity for frequency doubler and a cavity for SPDC via a Pound-Drever-Hall (PDH) locking method [27]. Most of laser is sent to a frequency doubler for generating 397.5 nm ultraviolet (UV) laser which is used to pump the nonlinear crystal to generate SPDC photon pairs (795 nm), and a small fraction of laser is split as the locking beam of cavity for SPDC.

The frequency doubler is in a standing-wave cavity configuration. A 10 mm long type-I PPKTP crystal is used for second-harmonic generation. By pumping 450 mW of light at 795 nm, we obtain 100 mW of 397.5 nm UV light. The cavity for SPDC is also in a standing-wave cavity configuration composed of two concave mirrors with curvature radius 50 mm. The cavity is near concentric in which the distance between two cavity mirrors is 92 mm. The input coupler mirror is high-reflection coated at 795 nm ( R＞99.9％) and anti-reflection (AR) coated at 397.5 nm. The output coupler mirror has a transmittance of 5% at 795 nm and high reflectance at 397.5 nm. A 10 mm long type-Ⅱ PPKTP crystal is used as the down-converter. Under the condition of phase matching, an H polarized UV pump photon down-converts to a near-infrared photon pair (795 nm) with one polarized in H and the other in V. The facets of the PPKTP are AR coated at 795 nm and 397.5 nm to minimize losses within the cavity. So the SPDC photons are resonant with the cavity and the UV pump light interacts twice with the PPKTP in the cavity. The purpose of UV light interacting twice is to facilitate the mode-matching between UV light beam and the cavity. The temperature of the type-Ⅱ PPKTP crystal is controlled by a high performance temperature controller with precision of 0.002℃. An isolator is inserted between the cavity for SPDC and frequency doubler to prevent the reflection of the cavity for SPDC from feeding back to the frequency doubler. We use a dichroic mirror DM1 to combine the locking light with the UV light before the cavity. After the cavity, DM2 is used to separate the residual UV light from the generated down-converted near-infrared photons. The near-infrared light transmits and the UV light reflects on DM2. Due to imperfect coating of DM2, there is a small fraction of near-infrared light in the reflective beam. The near-infrared light is extracted and detected by detector D1 which is used to observe the transmitted signal of cavity and lock the cavity. The cavity is locked at resonance using locking beam via PDH method, and the modulation frequency is 20 MHz. The frequency of degenerate mode ($\omega_0$) of the SPDC photons is as same as that of the locking beam. By changing the temperature of type-II PPKTP, the double resonance of H and V modes in the cavity for SPDC is realized. The temperature of PPKTP for double resonance is 40.356℃. Besides, two mechanical switches are used to alternate the periods of cavity locking and signal detection to avoid damaging the single photon detectors (SPDs) and eliminate the background count. The locking repetition rate is 20 Hz.

Due to the large phase-matching bandwidth of SPDC process (300 GHz), there are numerous non-degenerate correlated photon pairs together with degenerate photon pair. The quantum state from the cavity output can be expressed as

$$|\psi\rangle = \sqrt{\chi_0}|\omega_0\rangle_H|\omega_0\rangle_V$$
$$+ \sum_{m=1}^{N=100} \frac{\sqrt{\chi_m}}{2}(|\omega_0+m\Omega_H\rangle_H|\omega_0-m\Omega_H\rangle_V$$
$$+ |\omega_0-m\Omega_H\rangle_H|\omega_0+m\Omega_H\rangle_V \qquad (1)$$
$$+ |\omega_0+m\Omega_V\rangle_H|\omega_0-m\Omega_V\rangle_V$$
$$+ |\omega_0-m\Omega_V\rangle_H|\omega_0+m\Omega_V\rangle_V)$$

with $\frac{\chi_m}{\chi_0} = \frac{4}{1+\frac{4F^2}{\pi^2}\sin^2\frac{m\Delta\Omega}{\Omega}\pi}$, where $\Omega_H$ and $\Omega_V$ are the free spectrum ranges (FSR) for H and V modes, respectively. The measured values of $\Omega_H$ and $\Omega_V$ are 1468 MHz and 1481 MHz, with a difference $\Delta\Omega = \Omega_H - \Omega_V = 13\ MHz$. $F \approx 100$ is the finesse of the cavity. N is determined by the phase-matching bandwidth and the FSR of the cavity. In Eq. (1) only single photon-pair events are considered, and the vacuum states and multiple photon-pairs events are omitted. The first term of the right side of Eq. (1) is the expected degenerate pair output. The following four terms in the summation correspond to the case that one photon is resonant with the cavity while the other is not. The ratio between the summation of these non-degenerate modes to the degenerate mode is $\mu = \sum_{m=1}^{100}\frac{\chi_m}{\chi_0} = 1.87$. To filter the non-degenerate photon pairs, we employ five Fabry-Perot etalons, including two 5.4 mm-long, two 7.5 mm-long etalons and one 2.1 mm-long etalons. The reflectivities of two surfaces for all etalons are 90% and the measured finesses of them are ~25. The transmitted frequencies of these etalons are tuned to the frequency $\omega_0$ of degenerate photons by controlling their temperatures. Those etalons can keep stable operation in 24 hours. The total transmittance of the five etalons is ~40%. After the filters, the ratio $\mu$ is changed from 1.87 to $5.4 \times 10^{-6}$.

The degenerate single-mode photon pair is split into two parts with a 50/50 NPBS. Each photon of the photon pair has same probability to be transmitted or reflected at the NPBS. The reflected and transmitted photons are further coupled into two multimode fibers and detected by two SPDs (Perkin-Elmer SPCM-AQRH-15-FC) respectively. The signals from the SPDs are acquired by a four-channel event time digitizer (FAST Com Tec P7888) and analyzed by a computer. If two photons are detected by the two SPDs at the same time, respectively, the two photons are projected into entanglement state: $|\varphi\rangle = (1/\sqrt{2})(|H\rangle|V\rangle - e^{i\alpha}|V\rangle|H\rangle)$ [21], where $\alpha$ is the phase shift between $|H\rangle$ and $|V\rangle$ modes on the NPBS. In order to generate the maximum entanglement state $|\varphi\rangle = (1/\sqrt{2})(|H\rangle|V\rangle - |V\rangle|H\rangle)$, the phase shift is needed to be compensated. We add a phase compensation unit including two quarter wave plates (QWP) and one half wave plate (HWP) in one output path to compensate the phase shift.

**3. Results and discussion**

We firstly measured the generation rate of entanglement photon pairs as a function of the UV pump power. The measured coincidence counts are shown in Fig. 2. The data is fitted with a linear function which is good agreed with experimental data. We get a maximal generation rate of 1434 s$^{-1}$ at the pump power of 32 mW. By improving the efficiency of the filters and employing longer nonlinear crystal, the generation of entangled photon pairs can be increased further.

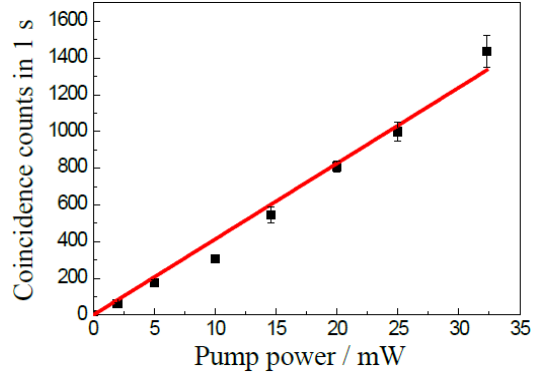

**Fig.2 (color online)** Coincidence counts in one second as a function of UV pump power. The black square dots are experimental data. The red line is the linear fitting.

The polarization correlation curves are measurement at 7 mW pump power. At a fixed polarization angle of photon A ($\theta_A$), coincidence counts between SPD1 and SPD2 over a 1 s interval are taken at different polarization angles of photon B ($\theta_B$). Fig.3 shows the coincidence counts for $\theta_A = 0$ (circle dots) and $\theta_A = -\pi/4$ (square dots), and their sinusoidal fits. We used the Monte-Carlo method to calculate the visibilities of two interference fringes. The visibilities are $0.978 \pm 0.015$ and $0.966 \pm 0.022$ for $\theta_A = 0$ and $\theta_A = -\pi/4$, respectively.

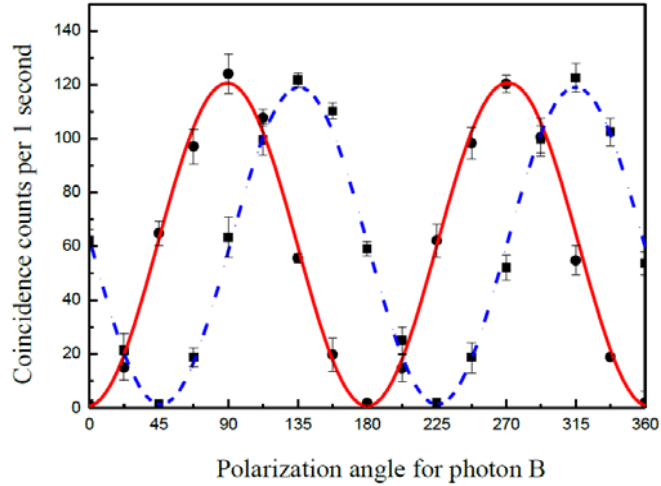

**Fig. 3 (color online)** Coincidence counts for $\theta_A = 0$ (black circle dots) and $\theta_A = -\pi/4$ (black square dots) as a function of $\theta_B$. The red (solid) and blue (dashed) curves are the sinusoidal fits to the experimental dates. The pump power is 7 mW.

It should be noted that the visibilities decrease with increasing pump power. The visibilities decrease to 92% at 20 mW pump power. The possible reason is that the large pump power causes the increase of multi-photon occurrence probability. We measure the Bell parameter S for the polarization-entangled photons at 7 mW pump power. In the canonical settings $\theta_A = 0$, $\theta'_A = \pi/4$, $\theta_B = \pi/8$, $\theta'_B = 3\pi/8$, we obtained an $S$ value $S = 2.73 \pm 0.04$ which violates the Bell-CHSH inequality by ~18 standard deviations.

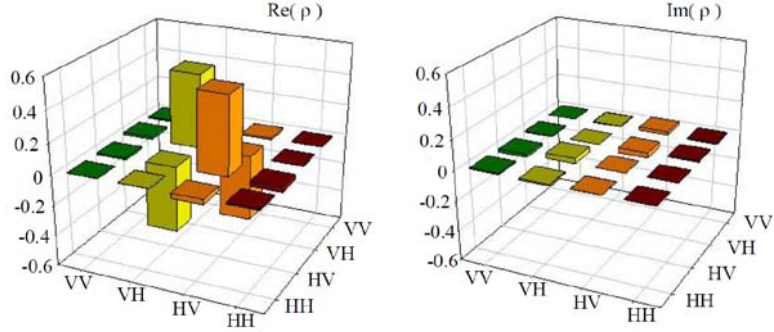

**Fig. 4 (color online)** The Real (a) and imaginary (b) parts of the density matrix of the generated entanglement photon pairs.

In order to further quantify the quantum properties of produced entanglement source, we also perform a quantum state tomography. In the process, 16 coincidence counts of independent projection states are needed to determine the polarization state density matrix. We insert an additional QWP in front of each polarizer, so the photon A and B can be projected into arbitrary polarization state by rotating the angle of the QWPs and HWPs. Using the maximum likelihood estimation method[28-29] the density matrix is reconstructed as follows:

$$\rho = \begin{pmatrix} 0.0104 & -0.0323 - 0.0012i & -0.0012 - 0.0082i & -0.0019 + 0.0073i \\ -0.0323 + 0.0012i & 0.5055 & -0.4269 + 0.0331i & -0.0113 - 0.0199i \\ -0.0012 + 0.0082i & -0.4269 - 0.0331i & 0.4762 & 0.0162 - 0.0047i \\ -0.0019 - 0.0073i & -0.0113 + 0.0199i & 0.0162 + 0.0047i & 0.008 \end{pmatrix}$$

Fig. 4 (a) and (b) show the real and imaginary parts of density matrix, respectively. By using the formula of $F = \langle \varphi | \rho | \varphi \rangle$, we calculate the fidelity of reconstructed density matrix by comparing with the maximally entangled state $|\varphi\rangle = (1/\sqrt{2})(|H\rangle|V\rangle - |V\rangle|H\rangle)$, which amounts to 95.2% ± 0.8%.

The time correlation function between the generated photon pairs is measured to acquire the linewidth of the entanglement source. In the experiment, the detected signal of SPD1 is sent to the start channel of P7888, and the detected signal of SPD2 is sent to the stop channel of P7888. The resolution time of P7888 is set to 1ns. By running the measurement 10000 times, we statistics the counts in each time delay, and get the curve of counts vs time delay, as shown in Fig. 5. The FWHM of correlation time is 14.5 ns. We use the function $\exp(-2\pi \Delta \nu |t|)$ to fit the experimental data, the best fit shows that the linewidth ($\Delta \nu$) is about 15 MHz, which matches the typical atomic memory bandwidth.

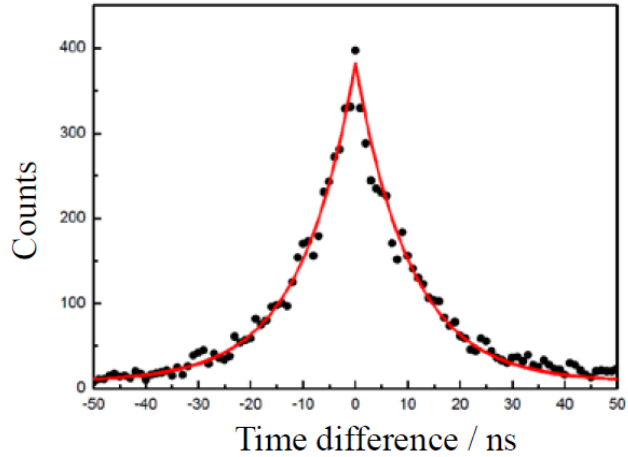

**Fig. 5 (color online)** The measured time correlation between generated photon pairs. A function of $\exp(-2\pi \Delta \nu |t|)$ is used to fit the experimental data (black dots). The red curve is the fitted result.

In Fig. 5, as the event time digitizer P7888 have a resolution time of 1 ns which is large than the cavity round-trip time of 670 ps, the time correlation measurement cannot prove our source single-mode output. To ensure the single-mode property of the generated photons, we measure its coherence length through a single-photon interference experiment [30-31] by using an Mach-Zehnder (M-Z) interferometer. The setup of M-Z interferometer is shown in Fig. 6(a). The $|H\rangle$ mode of the filtered photon pairs is sent to a fiber beam splitter which separates the field from input port into two equal parts at output1 and output2 ports. The two fields from output1 and output2 ports are combined with a 50/50 NPBS to form an M-Z interferometer. To scan the interference fringes, a reflective mirror in one arm is attached with a piozoelectric transducer for fine tuning. With a classical light fed into the interferometer, a interference visibility of 0.97 is observed, showing the good quality of interferometer. To change the path difference between two arms, we cut the fiber L2 and connect the fibers with different lengths between two end-points. The measured visibilities versus the path difference between two interfering beams are shown in Fig. 6(b). The average interference visibility is 95% when the two arms have same lengths. The visibility declines with the increase of the path difference. When the path difference is 4.5m, the visibility is still 75%. The coherence length is measured to be about 20 m, which is consistent with the result from the time correlation measurement in the relation of $\Delta L = \frac{\upsilon}{\Delta \upsilon} \lambda$. While for a multimode source, determined by the phase-matching condition, the coherence length is usually less than several mm. The measured coherence length provides good evidence that the generated photons are single-mode.

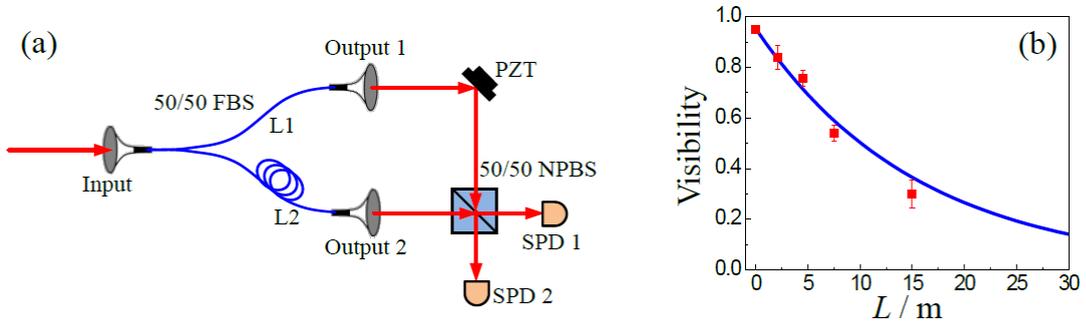

**Fig. 6 (color online)** (a) The setup of M-Z interferometer. 50/50 FBS: 50/50 fiber beam splitter; PZT: Piozoelectric transducer; 50/50 NPBS: 50/50 Non-polarizing beam splitter; SPD1-2: Single photon detectors. (b) The measured visibilities of the M-Z interferometer versus its optical path difference. The data are fitted with a function of $V = V_0 \, e^{-\frac{L}{L_0}}$. Error bars represent standard deviation.

## 4. Conclusion

We report an experimental preparation of a narrow-band polarization-entangled photon pairs through cavity-enhanced SPDC. The wavelength of entangled source is 795 nm, corresponding to $D_1$ line of Rb atoms. The single-mode photon pairs are selected by using multiple etalons as filters. The measured linewidth of entanglement source is 15 MHz which matches the typical atomic memory bandwidth. The measured bell parameter for the polarization-entangled photons S=2.73±0.04 which violates the Bell-CHSH inequality by ~18 standard deviations. By performing quantum state tomography, we obtain a fidelity of 95% between the generated entangled source and a maximally entangled state. The coherent length of entangled photons is measured to be about 20 m via an M-Z interferometer. The presented entanglement source could be utilized in quantum communication and quantum computing based on quantum memories in atomic ensemble.

**Acknowledgment**: This work was supported by the National Basic Research Program of China (Grant No. 2016YFA0301402), the National Natural Science Foundation of China (Grant Nos. 11475109, 11274211, 60821004), and the Program for Sanjin Scholars of Shanxi Province of China.

# References

[1] Duan L M, Lukin M D, Cirac J I and Zoller P 2001 Nature 414: 413-418

[2] Sangouard N, Simon C, Riedmatten H and Gisin N 2011 Rev Mod Phys 83, 33

[3] Knill E, Laflamme R and Milburn G J 2001 Nature 409:46-52

[4] Kok P, Munro W J, Nemoto K, Ralph T C, Dowling J P and Milburn G J 2007 Rev Mod Phys 79:135-174.

[5] Pan J W, Chen Z B, Lu C Y, Weinfurter H, Zeilinger A and Żukowski M 2012 Rev Mod Phys 84:777-838

[6] Cao D Y, Liu B H, Wang Z, Huang Y F, Li C F and Guo G C 2015 Sci Bull 60:1128-1132.

[7] Burnham DC and Weinberg D L 1970 Phys Rev Lett 25: 84–87.

[8] Kwiat P G, Mattle K, Weinfurter H, Zeilinger A, Sergienko A V and Shih Y 1995 Phys Rev Lett 75: 4337-4340.

[9] Kwiat P G, Waks E, White A G and Eberhard P H 1999 Phys Rev A 60: R773-R776.

[10] König F, Mason E J, Wong F N C and Albota M A 2005 Phys Rev A 71: 033805.

[11] Clausen C, Usmani I, Bussières F, Sangouard N, Afzelius M, Riedmatten H and Gisin N 2011 Nature 469: 508.

[12] Saglamyurek E, Sinclair N, Jin J, Slater J A, Oblak D, Bussiere F, George M, Ricken R, Sohler W and Tittel W 2011 Nature 469:512.

[13] Wu Y L, Li S J, Ge W, Xu Z X, Tian L and Wang H 2016 Sci Bull 61, 302.

[14] Akiba K, Kashiwagi K, Arikawa M, Kozuma M 2009 New J Phys 11, 013049.

[15] Reim K F, Nunn J, Lorenz V O, Sussman B J, Lee1 K C, Langford N K, Jaksch D and Walmsley I A et al 2010 Nat Photon 4: 218–221.

[16] Chen Y A, Chen S, Yuan Z S, Zhao Bo, Chuu C S, Schmiedmayer J and Pan J W 2008 Nat Phys 4: 103-107.

[17] Lvovsky AI, Sanders BC, Tittel W 2009 Nat Photon. 3:706-714.

[18] England DG, Michelberger PS, Champion TFM, Reim K F, Lee K C, Sprague M R , Jin X M, Langford N K, Kolthammer W S, Nunn J and Walmsley I A 2012 Phys B. 45, 124008.

[19] Ou Z Y, Lu Y J 1999 Phys Rev Lett 83:2556-2559.

[20] Wang H B, Horikiri T, Kobayashi T 2004 Phys Rev A 70, 043804.

[21] Kuklewicz C E, Wong F N C and Shapiro J H 2006 Phys Rev Lett 97, 223601.

[22] Scholz M, Wolfgramm F, Herzog U and Benson O 2007 Appl Phys Lett 91, 191104.

[23] Wang F Y, Shi B S and Guo G C 2010 Opt Commun 283:2974-2977.

[24] Morin O, Auria V D, Fabre C and Laurat J 2012 Opt Lett 37: 3738-3740.

[25] Bao XH, Qian Y, Yang J, Zhang H, Chen Z B, Yang T and Pan J W 2008 Phys Rev Lett 101, 190501.

[26] Zhang H, Jin X M, Yang J, Dai H N, Yang S J, Zhao T M, Rui J, He Y, Jiang X, Yang F, Pan G S, Yuan Z S, Deng Y J, Chen Z B, Bao X H, Chen S, Zhao B and Pan J W 2011 Nat Photon. 5, 628.

[27] Drever R W P, Hall J L, Kowaiski F V, Hough J, Ford G M, Munley A J and Ward H 1983 Appl. Phys. B 31, 97.

[28] White A G, James D F V, Eberhard P H and Kwiat P G 1999 Phys Rev Lett 83: 3103-3107.

[29] James D F V, Kwiat P G, Munro W J and White A G 2001 Phys Rev A 64, 052312

[30] Wang J, Lv P Y J, Cui J M, Liu B H, Tang J S, Huang Y F, Li C F and Guo G C 2015 Phys. Rev. Applied 4, 064011

[31] Wang F Y, Shi B S, Zhai C and Guo G C 2010 Journal of Modern Optics, 57, 330.